\documentclass[12pt]{article}

\begin{document}

\title{Nonlinear Amplitude Maxwell-Dirac
Equations. Optical Leptons}

\author{Lubomir M. Kovachev \\
Institute of Electronics, Bulgarian Academy of Sciences,\\
Tzarigradcko shossee 72,1784 Sofia, Bulgaria}

\maketitle

\abstract{ We apply the method of slowly-varying amplitudes of the
electrical and magnet fields to integro-differential system of
nonlinear Maxwell equations. The equations are reduced to system
of differential Nonlinear Maxwell amplitude Equations (NME).
Different orders of dispersion of the linear and nonlinear
susceptibility can be estimated. This method allow us to
investigate also the optical pulses with time duration equal or
shorter to the relaxation time of the media. The electric and
magnetic fields are presented as sums of circular and linear
components. Thus, NME is written  as a set of Nonlinear Dirac
Equations (NDE). Exact solutions of NDE with classical orbital
momentum $\ell = 1$ and opposite directions of the spin (opposite
charge) $j = \pm1 / 2$ are obtained. The possible generalization
of NME to higher number of optical components and higher number of
$\ell$ and $j$ is discussed. Two kind of Kerr type media: with and
without linear dispersion of the electric and the magnet
susceptibility are consider. The vortex solutions in case of media
with dispersion admit finite energy while the solutions in case of
a media without dispersion admit infinite energy.
}

\section{Introduction}

In the last four decades there has been considerable interest in
the nonlinear generalizations of the quantum field equations
\cite{IV,MAKH,RAJ} and in the possibility of obtaining exact
stationary solitary solutions of the field equations
\cite{BAR,FUSH}. As a rule, the nonlinearity has been introduced
in an {\it ad-hoc} fashion in the Klein-Gordon equation and also
for all four spinor components of the Dirac equations. For the
usual case of a cubic nonlinearity, exact localized solutions are
not found. Our present work, reported in this paper, shows that
{\it the optical analogy} of nonlinear Dirac equations leads to a
nonlinear part {\it only the first} coupled equation of Nonlinear
Dirac Equations. This result allows to solve the NDE by separation
of variables and to obtain solutions representing optical vortices
with classical momenta one and spin one-half. The initial
investigation of optical vortices began with a scalar theory,
based on the well-known 2D+1 paraxial Nonlinear Schrodinger
equation (NSE) \cite{SW,ROZ}. The existence of optical vortices
was predicted in the self-focusing regime, but as it was shown in
many papers, the solutions obtained are modulationally unstable.
In spite of this, various interactions (attraction, repulsion,
fusion) have been observed. The scalar paraxial approximation is
valid for slowly varying amplitudes of the electrical field in
weakly nonlinear media. As it was pointed out in
Refs.\cite{FF,JMN} this theory is not valid for a very intense
narrow pulses. The first generalization of the scalar paraxial
theory of optical vortices is based on investigation of so called
spatio-temporal scalar evolution equations \cite{SIL,DIN,EE}.  For
all cited theories no exact solutions have been found, but
numerical and energy momentum techniques are used. The existence
of exact stable vortex solutions of these types of nonlinear
equations was finally discovered with the vector generalization of
the 3D+1 Nonlinear Schrodinger equation \cite{AK}. It also have
been shown numerically that these vortices are stable at distances
comparable to those where localized solutions of the one component
scalar equation self-focusing rapidly. All of these exact vortices
are a combination of linearly-polarized components and have spin
$\ell = 1$. To extend the theory to vortices with spin $j = 1 / 2$
we return to an analogy between the Maxwell and Dirac equations .
As it is shown in Ref. \cite{DAC} this analogy is possible only if
the electrical and magnet components are represented as a sum of
linear and circularly polarized components.

\section{Maxwell's equations with non-stationary linear and nonlinear polarization}

Consider the Maxwell's equations in the next two cases:

$1$. A source-free medium with non-stationary linear and nonlinear
electric polarization and non-stationary magnetic polarization
(the case with dispersion).

$2$. A source-free medium with stationary linear and nonlinear
electric polarization and  stationary magnetic polarization (the
case without dispersion).

For these cases, the Maxwell's equations can be written:

\begin{eqnarray}
\label{eq1} \nabla \times \vec E = -\frac{1}{c}\frac{\partial\vec
B}{\partial t},
\end{eqnarray}

\begin{eqnarray}
\label{eq2} \nabla \times \vec H = \frac{1}{c}\frac{\partial \vec
D}{\partial t},
\end{eqnarray}

\begin{eqnarray}
\label{eq3} \nabla \cdot \vec D = 0,
\end{eqnarray}

\begin{eqnarray}
\label{eq4} \nabla \cdot \vec B = \nabla \cdot \vec H = 0,
\end{eqnarray}

\begin{eqnarray}
\label{eq5} \vec D = \vec P^{lin} + 4\pi \vec P_{n\ell}.
\end{eqnarray}
The linear magnetic polarization corresponds is [18,19]:

\begin{eqnarray}
\label{eq7} \vec B = \vec H + 4\pi \vec M_{lin},
\end{eqnarray}
where $\vec E$ and $\vec H$ are the electric and magnetic
intensity fields, $\vec D$ and $\vec B$ are the electric and
magnetic induction fields, $\vec P^{lin}$, $\vec P_{n\ell} $ are
the linear and nonlinear polarization of the medium respectively
and $\vec M_{lin}$ is the linear magnetic polarization. The
magnetic polarization (magnetization) $\vec {M}_{lin}$ is written
as the product of the linear magnetic susceptibility $\eta^{(1)}$
and the magnetic field $\vec H$.  The nonstationary linear
electric polarization can be written as:

\begin{eqnarray}
\label{eq8} \vec P^{lin} = \int\limits_{-\infty}^{t}
\left(\delta(t-\tau) + 4\pi\chi^{\left(1\right)}\left(t -
\tau\right)\right)
\vec E(\tau,x,y,z)d\tau \nonumber\\
=\int\limits_{-\infty}^{t}
 \varepsilon _0\left(t - \tau\right)\vec
E\left(\tau, x, y, z\right)d\tau,
\end{eqnarray}
where $\chi ^{(1)}$ and $\varepsilon_0$ are the linear electric
susceptibility and the dielectric constant respectively. Similar
expression describes the dependence of $\vec {B}$ on $\vec {H}$ in
the case of nonstationary linear magnetic polarization \cite{LAN}:

\begin{eqnarray}
\label{eq9} \vec B = \int\limits_{-\infty}^{t}
 \left(\delta(t-\tau) +
4\pi\eta^{\left(1\right)}\left(t - \tau\right)\right)
\vec H(\tau ,x,y,z)d\tau \nonumber\\
=\int\limits_{-\infty}^{t}
 \mu _0\left(t - \tau\right)\vec
H\left(\tau, x, y, z\right)d\tau,
\end{eqnarray}
where $\eta ^{(1)}$ and $\mu_0$ are the linear magnetic
susceptibility and magnetic permeability respectively. The
magnetic susceptibility of the main part of the dielectrics ranges
from $10^{-6}-10^{-4}$ and usually decreases with the increasing
of the frequency. In the following, we will study such media with
nonstationary cubic nonlinear polarization, where the nonlinear
polarization in the case of one carrying frequency can be
expressed as:

\begin{eqnarray}
\label{pnl} \vec P_{nl}^{(3)} = \frac{3}{4}
\int\limits_{-\infty}^{t}\int\limits_{-\infty}^{t}\int\limits_{-\infty}^{t}
\chi^{\left(3\right)}
\left(t-\tau_1,t-\tau_2,t-\tau_3\right)\nonumber\\
\times\vec{E}(\tau_1,x,y,z)\vec E^{*}(\tau_2,x,y,z)\vec
E(\tau_3,x,y,z) d\tau_1d\tau_2d\tau_3.
\end{eqnarray}
The causality request the next conditions on the response
functions:

\begin{eqnarray}
\label {caus} \varepsilon (t-\tau)=0;\ \chi^{\left(3\right)}
\left(t-\tau_1,t-\tau_2,t-\tau_3\right)=0, \nonumber\\
 t-\tau<0;\ t-\tau_i<0;\ i=1,2,3.
\end {eqnarray}
That why we can prolonged the upper integral boundary to infinity
and to use standard Fourier presentation \cite{MN}:

\begin{eqnarray}
\label {intcs} \int\limits_{-\infty}^{t}
{\varepsilon_0(\tau-t)\exp{(i\omega\tau)}d\tau}=\int\limits_{-\infty}^{+\infty}
{\varepsilon_0(\tau-t)\exp{(i\omega\tau)}d\tau}\\
\int\limits_{-\infty}^{t}\int\limits_{-\infty}^{t}\int\limits_{-\infty}^{t}
\chi^{\left(3\right)} \left(t-\tau_1,t-\tau_2,t-\tau_3\right)
d\tau_1d\tau_2d\tau_3=\nonumber\\
\int\limits_{-\infty}^{+\infty}\int\limits_{-\infty}^{+\infty}\int\limits_{-\infty}^{+\infty}
\chi^{\left(3\right)} \left(t-\tau_1,t-\tau_2,t-\tau_3\right)
d\tau_1d\tau_2d\tau_3.
\end {eqnarray}

The spectral presentation of linear optical susceptibility
$\hat{\varepsilon_{0}}(\omega)$ is connected to the nonstationary
optical response function by the next Fourier transform:

\begin{eqnarray}
\label{eq17} \hat{\varepsilon}_0(\omega)\exp{(i\omega t)}
=\int\limits_{-\infty}^{+\infty}
{\varepsilon_0(t-\tau)\exp{(i\omega\tau)}d\tau},
\end{eqnarray}
Similar expression for the spectral presentation of the
non-stationary magnetic response $\hat{\mu _ 0}(\omega)$:

\begin{eqnarray}
\label{eq18} \hat{\mu}_0(\omega)\exp{(-i\omega t)}
=\int\limits_{-\infty}^{+\infty}
{\mu_0(t-\tau)\exp{(-i\omega\tau)}d\tau}.
\end{eqnarray}
and nonlinear optical susceptibility $\chi^{(3)}$

\begin{eqnarray}
\label{chi3} \hat{\chi}^{(3)}=
\int\limits_{-\infty}^{+\infty}\int\limits_{-\infty}^{+\infty}
\int\limits_{-\infty}^{+\infty} \chi^{\left(3\right)}
\left(t-\tau_1,t- \tau_2,t-\tau_3\right)\nonumber\\
\times\exp{\left(i\left(\omega\left(\tau_1+\tau_2+\tau_3\right)\right)\right)}
d\tau_1d\tau_2d\tau_3
\end{eqnarray}
can be written. It is important to point here the remark of
Akhmanov at all in  \cite{AVC}, that such nonstationary
representation applied to slowly varying amplitudes of electrical
and magnet fields is valid as well when the optical pulse duration
of the pulses $t_0$ is greater than the characteristic response
time of the media $\tau_0$ ($t_0>>\tau_0$), as when the time
duration of the pulses are equal or less than the time response of
the media ($t_0\leq\tau_0$). We will discuss this possibility in
the process of deriving of the amplitude equations.

\section {Deriving of the amplitude equations}

In this section we derive the slowly varying amplitude
approximation in the standard way, as it was done in \cite{KAR},
\cite{MN}. For the  case of Maxwell equations with linear and
nonlinear dispersion (\ref{eq1})-(\ref{eq7}) we define the
electric and magnetic field amplitudes with the relations:

\begin{eqnarray}
\label{eq12} \vec E(x,y,z,t)=\vec
A(x,y,z,t)\exp{\left(i\left(\omega_0t-g(x,y,z)\right)\right)},
\end{eqnarray}

\begin{eqnarray}
\label{eq13} \vec H(x,y,z,t)=\vec
C(x,y,z,t)\exp{\left(-i\left(\omega_0t-q(x,y,z)\right)\right)},
\end{eqnarray}
where $\vec A$, $\vec C$, $\omega_0$,$g$ and $q$ are the
amplitudes of the electric and magnetic fields, the optical
frequency and the real spatial phase functions respectively. In
case of monochromatic and quasi-monochromatic fields the Stokes
parameters can be constructed from transverse components of the
wave field \cite{BORN,BERG}. This leads to two component vector
fields in a plane, transverse to the direction of propagation. For
electromagnetic fields with spectral bandwidth (our case) the two
dimensional coherency tensor cannot be used and the Stokes
parameters cannot be found directly. As it was shown by T.Carozzi
and all in \cite{BERG}, using high order of symmetry (SU(3)), in
this case six independent Stokes parameters can be found. This
corresponds to three component vector field. Here is investigated
this case. The increasing of the spectral bandwidth of the vector
wave, increases also the depolarization term (component, normal to
the standard Stokes coherent polarization plane).

We use the Fourier representation of the response functions
(\ref{eq17})-(\ref{chi3}) and of the amplitude functions $\vec
{A}$ (and $\vec {C}$) to obtain next expressions for the first
derivatives in time  of the linear polarization, nonlinear
polarization and magnetic induction fields:

\begin{eqnarray}
\label{eq21} \frac{1}{c}\frac{\partial \vec
P^{lin}\left(x,y,z,t\right)}{\partial t} =
i\exp{\left(i\left(\omega_0t-g(x,y,z)\right)\right)}\nonumber\\
\times\int\limits_{-\infty}^{+\infty}
\frac{\omega\hat{\varepsilon}_0\left(\omega\right)}{c} \vec
A(x,y,z,\omega-\omega_0)
\exp\left(i(\omega-\omega_0)t\right)d\omega,
\end{eqnarray}

\begin{eqnarray}
\label{dpnl} \frac{4\pi}{c}\frac{\partial \vec
P^{nlin}\left(x,y,z,t\right)}{\partial t} =
i\exp{\left(i\left(\omega_0t-g(x,y,z)\right)\right)}\nonumber\\
\times\int\limits_{-\infty}^{+\infty}
\frac{3\pi\omega\hat{\chi}^{(3)}\left(\omega\right)}{c} |\vec
A(x,y,z,\omega-\omega_0)|^2 \vec
A(x,y,z,\omega-\omega_0)\nonumber\\
\times\exp\left(i(\omega-\omega_0)t\right)d\omega.
\end{eqnarray}

\begin{eqnarray}
\label{eq22} -\frac{1}{c}\frac{\partial \vec
B\left(x,y,z,t\right)}{\partial t} =
i\exp{\left(i\left(\omega_0t-q(x,y,z)\right)\right)}\nonumber\\
\times\int\limits_{-\infty}^{+\infty}
\frac{\omega\hat{\mu}_0\left(\omega\right)}{c} \vec
C(x,y,z,\omega-\omega_0)
\exp\left(-i(\omega-\omega_0)t\right)d\omega.
\end{eqnarray}
At this point we restrict the spectrum of the amplitude of
electrical and magnet fields by writing the wave vectors
$k_{1,nl,2}$ in a Taylor series:

\begin{eqnarray*}
 k_1\left(\omega\right)=
 \frac{\omega\hat{\varepsilon}_0\left(\omega\right)}{c}=k_1^0\left(\omega_0\right)+
\end{eqnarray*}

\begin{eqnarray*}
  + \frac{\partial\left(k_1\left(\omega_0\right)\right)}{\partial\omega}
\left(\omega-\omega_0\right) + \frac{1}{2}\frac{\partial
^2\left(k_1\left(\omega_0\right)\right)}
{\partial\omega^2}\left(\omega-\omega_0\right)^2 + ....=
\end{eqnarray*}

\begin{eqnarray}
\label{eq23} =k_1^0\left(\omega_0\right)+
\frac{1}{v_1}\left(\omega-\omega_0\right)
+\frac{1}{2}k_1^{"}\left(\omega-\omega_0\right)^2+ ....,
\end{eqnarray}

\begin{eqnarray*}
 k_{nl}\left(\omega\right)=
 \frac{3\pi\omega\hat{\chi}^{(3)}\left(\omega\right)}{c}=
\end{eqnarray*}

\begin{eqnarray*}
 =k_{nl}^0\left(\omega_0\right) +
\frac{\partial\left(k_{nl}\left(\omega_0\right)\right)}{\partial\omega}
\left(\omega-\omega_0\right) + ....=
\end{eqnarray*}

\begin{eqnarray}
\label{knl} =k_{nl}^0\left(\omega_0\right)+
\frac{1}{v_{nl}}\left(\omega-\omega_0\right) + ....,
\end{eqnarray}

\begin{eqnarray*}
k_2\left(\omega\right)=
\frac{\omega\hat{\mu}_0\left(\omega\right)}{c}=
k_2^0\left(\omega_{0}\right)
\end{eqnarray*}

\begin{eqnarray*}
+
\frac{\partial\left(k_2\left(\omega_0\right)\right)}{\partial\omega}
\left(\omega-\omega_0\right) + \frac{1}{2}\frac{\partial
^2\left(k_2\left(\omega_0\right)\right)} {\partial\omega
^2}\left(\omega-\omega_0\right)^2 + .... =
\end{eqnarray*}

\begin{eqnarray}
\label{eq24}
k_2^0\left(\omega_0\right)+
\frac{1}{v_2}\left(\omega-\omega_0\right)
+\frac{1}{2}k_2^{"}\left(\omega-\omega_0\right)^2 +....,
\end{eqnarray}
where $v_{i}$ and $k_{i}^{"};i = 1,nl,2$ have dimensions of group
velocity, and nonlinear addition to the group velocity and
dispersion respectively.  The nonlinear wave vector is expressed:

\begin{eqnarray}
k_{nl}^0=3\pi\omega_0\hat{\chi}^{(3)}(\omega_0)/c=
\frac{\omega_0\hat{\varepsilon}_0}{c}
\frac{3\pi\hat{\chi}^{(3)}(\omega_0)}{\hat{\varepsilon}_0}
=k_1n_2,
\end{eqnarray}
where

\begin{eqnarray}
n_2(\omega_0)=\frac{3\pi\hat{\chi}^{(3)}(\omega_0)}{\hat{\varepsilon}_0},
\end{eqnarray}
is the nonlinear refractive index. It is important to note here
that we do not use any approximation of the response function.
There is only one requirement of the spectral presentations
(\ref{chi3}),(\ref{eq17}) and (\ref{eq18}) of the response
functions: to admit first and second order derivatives in respect
to frequency ( to be smooth functions). The restriction is only in
respect of the relation between the main frequency $\omega_0$ and
time duration of the envelope functions $t_0$ determinate from the
relations (\ref{eq23}), (\ref{knl}) and (\ref{eq24}) (conditions
for slowly varying amplitudes). Putting eqn. (\ref{eq23}) in
(\ref{eq21}), (\ref{eq24}) in (\ref{eq22}) and (\ref{knl}) in
(\ref{dpnl}), and keeping in mind the expressions for time
derivatives of the spectral presentation of the amplitude
functions, the first derivatives (\ref{eq21}), (\ref{dpnl}) and
(\ref{eq22}) are presented in form:

\begin{eqnarray}
\label{eq25} \frac{1}{c}\frac{\partial \vec
P^{lin}\left(x,y,z,t\right)}{\partial t} = \left(ik_1^0\vec A +
\frac{1}{v_1}\frac{\partial \vec A}{\partial t}-
\frac{ik_1^{"}}{2}\frac{\partial ^{2}\vec A}{\partial
t^2}\right)\nonumber\\
\times\exp{\left(i\left(\omega_0t-g(x,y,z)\right)\right)},
\end{eqnarray}

\begin{eqnarray}
\label{apnl} \frac{4\pi}{c}\frac{\partial \vec
P^{nlin}\left(x,y,z,t\right)}{\partial t} =\left( ik_1^0n_2|\vec
A|^2 \vec A + \frac{1}{v_{nl}}\frac{\partial |\vec A|^2\vec
A}{\partial t}\right)\nonumber\\
\times\exp{\left(i\left(\omega_0t-g(x,y,z)\right)\right)},
\end{eqnarray}

\begin{eqnarray}
\label{eq26} -\frac{1}{c}\frac{\partial \vec
B\left(x,y,z,t\right)}{\partial t} = \left(ik_2^0\vec C +
\frac{1}{v_2}\frac{\partial \vec C}{\partial t}-
\frac{ik_2^{"}}{2}\frac{\partial^{2}\vec C}{\partial
t^2}\right)\nonumber\\
\times\exp{\left(i\left(\omega_0t-q(x,y,z)\right)\right)}.
\end{eqnarray}
Finally, from the Maxwell equations (\ref{eq1})-(\ref{eq7}), using
eq. (\ref{eq25}), eq. (\ref{apnl}) and eq.(\ref{eq26}), and using
in fact that $\nabla \cdot \vec {D} \approx \nabla \cdot \vec
{E}\approx 0$ \cite {AK}, we obtain the next system of Nonlinear
Maxwell vector amplitude Equations (NME):

\begin{eqnarray}
\label{eq28} \nabla \times \vec {A} = ik_2^0 \vec C -
\frac{1}{v_2}\frac{\partial \vec C}{\partial t} -
\frac{ik_2^{"}}{2}\frac{\partial ^2\vec C}{\partial t^2},
\end{eqnarray}

\begin{eqnarray}
\label{eq29} \nabla \times \vec {C} = ik_1^0 \vec A +
\frac{1}{v_1}\frac{\partial \vec A}{\partial t} -
\frac{ik_1^{"}}{2}\frac{\partial ^2\vec A}{\partial t^2} +
ik_1^0n_2 \left(\vec A\cdot\vec A^{\ast}\right)\vec A \nonumber\\
+ \left(\frac{n_2}{v_1}+k_1\frac{\partial
n_2}{\partial\omega}\right)\frac{\partial\left(\vec A\cdot\vec
A^{\ast}\right)\vec A} {\partial t},
\end{eqnarray}

\begin{eqnarray}
\label{eq30} \nabla \cdot \vec {A} = 0,
\end{eqnarray}

\begin{eqnarray}
\label{eq31} \nabla \cdot \vec {C} = 0,
\end{eqnarray}
if the gradient of the spatial phase functions $g$ and $q$
satisfied the relations:

\begin{eqnarray}
\label{phas1} \nabla g\times \vec {A} = 0,
\end{eqnarray}
and

\begin{eqnarray}
\label{phas2} \nabla q\times \vec {C} = 0.
\end{eqnarray}
The phase functions whichever satisfied
(\ref{phas1})-(\ref{phas2}) are determinate  in Section ($10$).

 We investigate the case, when our vector fields are
presented as a sum of circular and linear polarizing components:

\begin{eqnarray}
\label{pol} \vec {A} =\vec {A}_{lin} +\vec {A}_{cir},\\
\vec {C} =\vec {C}_{lin} +\vec {C}_{cir}.
\end{eqnarray}
The nonlinear polarization admit different nonlinear refractive
indexes in the case of linear and circular polarization
\cite{BOYD} ( $n_2^{lin}\neq n_2^{cir}$). We will include this
difference in the our rescaled equations, defining the next
rescaled dependant variables:

\begin{eqnarray}
\label{scal}
\vec {A} =A_{0}^{lin}\vec {A}_{lin}^{'} +A_0^{cir}\vec {A}_{cir}^{'},\\
\vec {C} =C_0(\vec {C}_{lin}^{'} +\vec {C}_{cir}^{'}),\\
(A_0^{lin})^2=\frac{n_2^{cir}}{n_2^{lin}}(A_0^{cir})^2,
\end{eqnarray}
and independent variables;

\begin{eqnarray} \label{eq32}
x = r_{0} x^{'};y = r_{0} y^{'};z = r_{0} z^{'},t = t_{0} t^{'}.
\end{eqnarray}
In additions the next constants are determinate:

\begin{eqnarray}
\label{eq33}
\begin{array}{l}
\alpha _{i} = k_{i}^{0} r_{0};\ \beta _{i} = k_{i}^{"} r_{0} /
2t_{0}^{2};\ \gamma_1 = r_0k_1n_2^{cir}\left|A_0\right|^2; \nonumber\\
 \gamma _{2} =n_2^{cir}\left|A_0\right|^2;\
 \gamma _{3} = v_{1}n_2^{cir}\left|A_0\right|^2/c;\ \nonumber\\
 \delta = v_1/v_2\le 1;\ v_1\approx r_0/t_0;\ i = 1,2.
 \end{array}
\end{eqnarray}
The NME (\ref{eq28})-(\ref{eq31})in rescaled variables are
transformed into the following (the primes have been omitted for
clarity):

\begin{eqnarray}
\label{eq34}
\nabla \times \vec {A} = i\alpha _{2} \vec {C} -
\delta\frac{\partial\vec C}{\partial t} -
i\beta_2\frac{\partial ^2\vec C}{\partial t^2},
\end{eqnarray}

\begin{eqnarray}
\label{eq35} \nabla \times \vec {C} = i\alpha _{1} \vec {A} +
\frac{\partial\vec A}{\partial t} - i\beta_1\frac{\partial^2\vec
A}{\partial t^2} + i\gamma_1\left(\vec A\cdot\vec
A^{\ast}\right)\vec A \nonumber\\+
\left(\gamma_2+\gamma_3\right)\frac{\partial\left(\left(\vec
A\cdot\vec A^{\ast}\right)\vec A\right)} {\partial t},
\end{eqnarray}

\begin{eqnarray}
\label{eq36} \nabla \cdot \vec {A} = 0,
\end{eqnarray}

\begin{eqnarray}
\label{eq37} \nabla \cdot \vec {C} = 0.
\end{eqnarray}
We consider the case of slowly varying amplitudes approximation
when the nonlinear constant is $\gamma_{1} = 1$. Then the
constants in fronts of last therm on right hand are $\gamma
_{2}\approx \gamma_3\approx 10^{-2} - 10^{-3}$. While the
constants $\alpha _{i}$ have typical values
$\alpha\approx10^2-10^3\left(\alpha_i\sim r_0k_i\right)$, the
dispersion  constants $\beta _{i} $ have very small values $\beta
_1\sim10^{-5}-10^{-6}$ for picosecond and sub-picosecond pulses in
the transparency region of nonlinear optical media. Neglecting the
small dispersion terms $\beta _{i}<<1$ and the last nonlinear term
for witch $\gamma _{2}\approx\gamma_3<<1$, the NME system
(\ref{eq34})-(\ref{eq37})can be rewritten:

\begin{eqnarray}
\label{eq38} \nabla \times\vec {A} = i\alpha_2\vec C -
\delta\frac{\partial\vec C}{\partial t},
\end{eqnarray}

\begin{eqnarray}
\label{eq39} \nabla\times\vec {C} = i\alpha_1\vec A +
\frac{\partial \vec A}{\partial t} + i\gamma_1\left(\vec
A\cdot\vec A^{\ast}\right)\vec A,
\end{eqnarray}

\begin{eqnarray}
\label{eq40} \nabla \cdot\vec {A} = 0,
\end{eqnarray}

\begin{eqnarray}
\label{eq41}
\nabla \cdot \vec {C} = 0.
\end{eqnarray}

\section{Dirac representation of NME}

To solve the NME (\ref{eq38})-(\ref{eq41}), we apply the method of separation
of variables. The slowly varying amplitude vector of the electric field
$\vec {A}$ and the magnetic field $\vec {C}$ are represented as:

\begin{eqnarray}
\label{eq42} \vec A\left(x,y,z,t\right)= \vec F\left(x,y,z\right)
\exp{\left(i\Delta\alpha t\right)},
\end{eqnarray}

\begin{eqnarray}
\label{eq43} \vec C\left(x,y,z,t\right)= \vec G\left(x,y,z\right)
\exp{\left(i\Delta\alpha t\right)}.
\end{eqnarray}
Substituting these forms into the NME (\ref{eq38})-(\ref{eq41}) we
obtain:

\begin{eqnarray}
\label{eq44} \nabla\times\vec {F} = - i\nu _2 \vec {G},
\end{eqnarray}

\begin{eqnarray}
\label{eq45} \nabla \times \vec {G} = i\nu_1\vec F +
i\gamma_1\left(\vec F\cdot \vec F^{\ast}\right)\vec F,
\end{eqnarray}

\begin{equation}
\label{eq46} \nabla \cdot \vec {F} = 0,
\end{equation}

\begin{equation}
\label{eq47}
\nabla \cdot \vec {G} = 0,
\end{equation}
where $\nu_1 = \alpha_1 + \Delta \alpha $; $\nu _2 = \delta
\Delta\alpha - \alpha _2 > 0$. In a Cartesian coordinate system,
the vector equations (\ref{eq44})-(\ref{eq47}) are reduced to a
scalar system of eight nonlinear wave equations. When the electric
and magnetic fields are represented as a sum of a linear
polarization component and a circular polarized one it is possible
to reduce eqns. (\ref{eq44}) - (\ref{eq47}) to a system of four
nonlinear equations. Substituting \cite {DAC}:

\begin{eqnarray}
\label{eq48} \psi _{1} = iF_l = iF_z,
\end{eqnarray}

\begin{eqnarray}
\label{eq49} \psi _{2} = F_{c} = iF_{x} - F_{y},
\end{eqnarray}

\begin{eqnarray}
\label{eq50} \psi_{3} = G_l=-G_{z},
\end{eqnarray}

\begin{eqnarray}
\label{eq51}
\psi_{4}=G_{c}=-G_{x} - iG_{y},
\end{eqnarray}
into the nonlinear system (\ref{eq44})-(\ref{eq47}), we obtain a stationary
nonlinear Dirac system of equations (NDE):

\begin{eqnarray}
\label{eq52} \left(\frac{\partial}{\partial
x}-i\frac{\partial}{\partial y}\right) \Psi_4 +
\frac{\partial}{\partial z}\Psi_3 = -
i\left(\nu_1+\gamma_1\sum\limits_{i =
1}^{2}{{\left|\Psi_i\right|}^2}\right) \Psi_1,
\end{eqnarray}

\begin{eqnarray}
\label{eq53} \left(\frac{\partial}{\partial
x}+i\frac{\partial}{\partial y}\right) \Psi_3 -
\frac{\partial}{\partial z}\Psi_4 = -
i\left(\nu_1+\gamma_1\sum\limits_{i =
1}^{2}{{\left|\Psi_i\right|}^2}\right) \Psi_2,
\end{eqnarray}

\begin{eqnarray}
\label{eq54} \left(\frac{\partial}{\partial x} -
i\frac{\partial}{\partial y}\right) \Psi_2 +
\frac{\partial}{\partial z}\Psi_1 = - i\nu_2\Psi_3,
\end{eqnarray}

\begin{eqnarray}
\label{eq55}
\left(\frac{\partial}{\partial x} + i\frac{\partial}{\partial y}\right)
\Psi_1 - \frac{\partial}{\partial z}\Psi_2 = -
i\nu_2\Psi_4.
\end{eqnarray}
This substitution allows to reduce the system of eight equations
(\ref{eq44})-(\ref{eq47}) to a system of four scalar complex
equations (\ref{eq52})-(\ref{eq55}). The system
(\ref{eq52})-(\ref{eq55}) is the optical analog of the nonlinear
Dirac equations (NDE). Note that the {\it optical} NDE are
significantly different from the NDE in the field theory. The
nonlinear part appears {\it only} in the first two coupled
equations of the system.

\section{ Vortex solutions with orbital momentum  l=1 and spin j=1/2}

The symmetries of the NDE (\ref{eq52})-(\ref{eq55}) are used to
obtain exact vortex solutions. The NDE (\ref{eq52})-(\ref{eq55})
have both spherical and spinor symmetry only in the case where the
nonlinear part does not manifest the angular dependence on the
radial variable
$\sum\limits_{i=1}^{2}{{\left|\Psi_i\left(r,\theta,\varphi\right)\right|}^2}
 = F\left(r\right)$.
This type of solution can be found using with the following
technique. Using Pauli matrices, we write the NDE system
(\ref{eq52})-(\ref{eq55}) as:

\begin{eqnarray}
\label{eq56}
\left(\vec{\sigma}\cdot\vec P\right)\phi =
-i\left(\nu_1+\gamma_1\sum\limits_{i=1}^{2}{{\left|\eta_i\right|}^2}\right)
\eta,
\end{eqnarray}

\begin{eqnarray}
\label{eq57}
\left(\vec{\sigma}\cdot\vec P\right)\eta = -i\nu_2\phi,
\end{eqnarray}
where

$\vec{\sigma} = \left[ {\left(
\begin{array}{*{20}c}
 {0} \hfill & {1} \hfill \\
 {1} \hfill & {0} \hfill \\
\end{array}
\right),\left(
\begin{array}{*{20}c}
 {0} \hfill & { - i} \hfill \\
 {i} \hfill & {0} \hfill \\
\end{array}
 \right),\left(
 \begin{array}{*{20}c}
 {1} \hfill & {0} \hfill \\
 {0} \hfill & { - 1} \hfill \\
 \end{array}\right)}\right]$,

are the Pauli matrices, $\vec P = \left(\frac{\partial}{\partial
x}, \frac{\partial}{\partial y},\frac{\partial}{\partial
z}\right)$

is the differential operator and

$\eta = \left(
\begin{array}{*{20}c}
 \Psi_1  \hfill \\
 \Psi_2  \hfill \\
\end{array}
\right)$;
$\phi = \left(
\begin{array}{*{20}c}
 \Psi_3  \hfill \\
 \Psi_4  \hfill \\
\end{array}
\right)$,

are the corresponding spinors. After substituting eqn.(\ref{eq57})
into eqn.(\ref{eq56}) we obtain:

\begin{equation}
\label{eq58}
\left(\vec{\sigma}\cdot\vec P\right)\left(\vec\sigma\cdot\vec P\right)\eta=
-\nu_2\left(\nu_1+
\gamma_1\sum\limits_{i=1}^{2}{{\left|\eta_i\right|}^2}\right)\eta.
\end{equation}
When there is no external electric or magnetic field, the operator
on the left-hand side of eqn. (\ref{eq58}) is  is the Laplacian
operator $\Delta $:

\begin{equation}
\label{eq59} P^2=\left(\vec{\sigma}\cdot\vec
P\right)\left(\vec{\sigma} \cdot\vec P\right) = \Delta.
\end{equation}
 From (\ref{eq58}) and (\ref{eq59}) we obtain:

\begin{equation}
\label{eq60} \nu_2\nu_1\eta+\nu_2\gamma_1\sum\limits_{i=1}^{2}
{{\left|\eta_i\right|}^2}\eta + \Delta \eta = 0.
\end{equation}
In case of spherical representation of the spinor equations
(\ref{eq58}), there are two possibilities, $l=0$ and $l=1$, that
will permit only a radial dependence of the nonlinear part:

\begin{equation}
\label{eq61}
\sum\limits_{i=1}^{2}{{\left|\eta_i\left(r,\theta,\varphi\right)\right|}^2}=
F\left(r\right).
\end{equation}
For the $l = 0$ case, the equations (\ref{eq60}) are transformed
to equations with radial parts:

\begin{equation}
\label{eq62}
\nu_2\nu_1\eta+\nu_2\gamma_1
\sum\limits_{i=1}^{2}{{\left|\eta_i\right|}^2}\eta+
\frac{\partial^2\eta}{\partial r^2} +
\frac{2}{r}\frac{\partial \eta}{\partial r} = 0.
\end{equation}
The scalar variant of these equations has been investigated in
many papers but exact localized solutions have not been found. In
the case $l = 1$ we look for spinors in the next two forms:

\begin{eqnarray}
\label{eq63}
\eta = \left(
\begin{array}{*{20}c}
\tilde{\eta}\left(r\right)\cos\left(\theta\right)
\hfill \\
\tilde{\eta}\left(r\right)\sin\left(\theta\right)
\exp{\left(i\varphi\right)}
\hfill \\
\end{array}
\right).
\end{eqnarray}
and

\begin{eqnarray}
\label{eq63a} \eta = \left(
\begin{array}{*{20}c}
\tilde{\eta}\left(r\right)\sin\left(\theta\right)
\exp{\left(-i\varphi\right)}
\hfill \\
-\tilde{\eta}\left(r\right)\cos\left(\theta\right)
\hfill \\
\end{array}
\right).
\end{eqnarray}
As it will be seen later this corresponds to two opposite
directions of the own orbital momentum $j=\pm1/2$ (opposite
charge). After substituting solutions (\ref{eq63})-(\ref{eq63a})
into equations (\ref{eq60}) the following equation describing the
radial dependence is obtained:

\begin{equation}
\label{eq64}
\nu_2\nu_1\tilde{\eta} + \nu_2\gamma_1
{\left|\tilde{\eta}\right|}^2\tilde{\eta}  +
\frac{\partial^2\tilde{\eta}}{\partial r^2} +
\frac{2}{r}\frac{\partial \tilde {\eta} }{\partial r} -
\frac{2}{r^2}\tilde{\eta}= 0.
\end{equation}
The angular parts are the standard spherical harmonics with $l =
1$. This system has exact vortex {\it de Broglie} soliton
solutions \cite{BAR} in the form:

\begin{equation}
\label{eq65} \tilde{\eta}\left(r\right) = \frac{\sqrt
2}{i}\frac{\exp{\left(i\sqrt{\nu_1\nu_2}r\right)}}{r},
\end{equation}
if $\nu_{2}\gamma_{1}=1$. The complete solutions for these two
cases are written:

\begin{eqnarray}
\label{eq66}
\eta = \left(
\begin{array}{*{20}c}
\frac{\sqrt 2}{i}\frac{\exp(i\sqrt{\nu_1\nu_2}r)}{r}
\cos\left(\theta\right)\\
\hfill \\
\frac{\sqrt 2}{i}\frac{\exp(i\sqrt{\nu_1\nu_2}r)}{r}
\sin\left(\theta\right)\exp{\left(i\varphi\right)}
\hfill \\
\end{array}
\right),
\end{eqnarray}

\begin{eqnarray}
\label{eq67}
\phi = \left(
\begin{array}{*{20}c}
\frac{\sqrt 2}{i\nu_2}
\left(\frac{i\sqrt{\nu_1\nu_2}\exp{(i\sqrt{\nu_1\nu_2}r)}}{r} +
\frac{\exp{i(\sqrt{\nu_1\nu_2}r)}}{r^2}\right)\\
\hfill \\
0
\hfill \\
\end{array}
\right),
\end{eqnarray}

and
\begin{eqnarray}
\label{eq66a} \eta = \left(
\begin{array}{*{20}c}
\frac{\sqrt 2}{i}\frac{\exp(i\sqrt{\nu_1\nu_2}r)}{r}
\sin\left(\theta\right)\exp{-\left(i\varphi\right)}\\
\hfill \\
\frac{-\sqrt 2}{i}\frac{\exp(i\sqrt{\nu_1\nu_2}r)}{r}
\cos\left(\theta\right)
\hfill \\
\end{array}
\right),
\end{eqnarray}

\begin{eqnarray}
\label{eq67a} \phi = \left(
\begin{array}{*{20}c}
0\\
\hfill \\
\frac{-\sqrt 2}{i\nu_2}
\left(\frac{i\sqrt{\nu_1\nu_2}\exp{(i\sqrt{\nu_1\nu_2}r)}}{r} +
\frac{\exp{i(\sqrt{\nu_1\nu_2}r)}}{r^2}\right)
\hfill \\
\end{array}
\right).
\end{eqnarray}
There is another, more direct way for separating the variables of
the spinor equations (\ref{eq56})-(\ref{eq57}). To illustrate
this, we represent the NDE (\ref{eq52})-(\ref{eq55}) in spherical
variables:

\begin{eqnarray}
\label{eq68}
\exp{(-i\varphi)}\left(\sin\theta\frac{\partial}{\partial r} +
\frac{\cos\theta}{r}\frac{\partial}{\partial\theta}-\frac{i}{r\sin\theta}
\frac{\partial}{\partial\varphi}\right)\Psi_4 \nonumber\\
+\left(\cos\theta\frac{\partial}{\partial r} -
\frac{\sin\theta}{r}\frac{\partial}{\partial\theta}\right)\Psi_3
=-i\left(\nu_1+\gamma_1\sum\limits_{i=1}^2{\left|\Psi_i\right|}^2\right)
\Psi_1,
\end{eqnarray}

\begin{eqnarray}
\label{eq69}
\exp{(i\varphi)}\left(\sin\theta\frac{\partial}{\partial r} +
\frac{\cos\theta}{r}\frac{\partial}{\partial\theta}+\frac{i}{r\sin\theta}
\frac{\partial}{\partial\varphi}\right)\Psi_3 \nonumber\\
-\left(\cos\theta\frac{\partial}{\partial r} -
\frac{\sin\theta}{r}\frac{\partial}{\partial\theta}\right)\Psi_4
=-i\left(\nu_1+\gamma_1\sum\limits_{i=1}^2{\left|\Psi_i\right|}^2\right)
\Psi_2,
\end{eqnarray}

\begin{eqnarray}
\label{eq70}
\exp{(-i\varphi)}\left(\sin\theta\frac{\partial}{\partial r} +
\frac{\cos\theta}{r}\frac{\partial}{\partial\theta}-\frac{i}{r\sin\theta}
\frac{\partial}{\partial\varphi}\right)\Psi_2 \nonumber\\
+\left(\cos\theta\frac{\partial}{\partial r} -
\frac{\sin\theta}{r}\frac{\partial}{\partial\theta}\right)\Psi_1
=-i\nu_2\Psi_3,
\end{eqnarray}

\begin{eqnarray}
\label{eq71}
\exp{(i\varphi)}\left(\sin\theta\frac{\partial}{\partial r} +
\frac{\cos\theta}{r}\frac{\partial}{\partial\theta}+\frac{i}{r\sin\theta}
\frac{\partial}{\partial\varphi}\right)\Psi_1 \nonumber\\
-\left(\cos\theta\frac{\partial}{\partial r} -
\frac{\sin\theta}{r}\frac{\partial}{\partial\theta}\right)\Psi_2
=-i\nu_2\Psi_4.
\end{eqnarray}
We make the following two {\it ansatzes} for solutions to the
system of nonlinear equations (\ref{eq68})-(\ref{eq71}):

\begin{eqnarray}
\label{eq72}
\begin{array}{l}
 \Psi _{1} = a\left( {r} \right)\cos \left( {\theta}  \right) \\
 \Psi _{2} = a\left( {r} \right)\sin \left( {\theta}  \right)e^{i\varphi}
\\
 \Psi _{3} = - ib\left(r\right) \\
 \Psi _{4} = 0. \\
 \end{array}
\end{eqnarray}
and

\begin{eqnarray}
\label{eq72a}
\begin{array}{l}
 \Psi _{1} = a\left( {r} \right)\sin \left( {\theta}  \right)e^{i\varphi} \\
 \Psi _{2} = -a\left( {r} \right)\cos \left( {\theta}  \right)\\
 \Psi _{3} = 0 \\
 \Psi _{4} = - ib\left(r\right). \\
 \end{array}
\end{eqnarray}
Put (\ref{eq72}) or  (\ref{eq72a}) in equations
(\ref{eq68})-(\ref{eq71}) we separate the variables. The following
system of equations describing the radial dependence of the
amplitudes are obtained:

\begin{equation}
\label{eq73} \frac{\partial a\left(r\right)}{\partial r} +
\frac{2}{r}a\left(r\right) = -\nu_2b\left(r\right),
\end{equation}

\begin{equation}
\label{eq74}
\frac{\partial b\left(r\right)}{\partial r}=
\nu_1a\left(r\right)+\gamma_1{\left|a\left(r\right)\right|}^2a\left(r\right).
\end{equation}
In the system of equations (\ref{eq73})-(\ref{eq74}) the
nonlinearity appears only in the radial part, while for the
angular part we have standard spherical spinors with spin $j=\pm
1$. Solving eqns.(\ref{eq73}) and (\ref{eq74}) we
straightforwardly show that when $\nu_{2} \gamma=1$ these
equations admit the localized {\it de Broglie} solitons of eqns.
(\ref{eq66})-(\ref{eq67}).

\section{ Hamiltonian representation of the NDE. First integrals for vortex solutions with spin
$j = \pm1/2$}

It is not difficult to show that for the NDE system of eqns. (\ref{eq52})-
(\ref{eq55}) a Hamiltonian of the form:

\begin{equation}
\label{eq80} H = \left(\vec{\sigma}\cdot\vec
P\right)+\sum\limits_{i=1}^{2} {{\left|\Psi_i\right|}^2},
\end{equation}
can be written. Using this, eqns. (\ref{eq52})-(\ref{eq55}) can be
rewritten in the form:

\begin{equation}
\label{eq81}
H\Psi = \varepsilon \Psi,
\end{equation}
where
$\varepsilon=\left(-i\nu_1,-i\nu_1,-i\nu_2,-i\nu_2\right)$
is the energy operator.Here we investigate the case where the nonlinear part
of the equation is represented as a number of spinors with a scalar sum that
depends only on the radial component.

\begin{equation}
\label{eq82}
\sum\limits_{i=1}^2{{\left|\Psi_i\left(r,\theta,\varphi\right)\right|}^2}
=F\left(r\right).
\end{equation}
We also introduce here the well known orbital momentum operator
$\vec {L}$, own orbital (spin) momentum $\vec {S}$, and the full
momentum $\vec {J}$, as well as:

\begin{equation}
\label{eq83}
\vec {L}=\vec {r}\times\vec {P}
\end{equation}

\begin{eqnarray}
\label{eq84}
\frac{1}{2}\vec{S}=\frac{1}{2}\left(
\begin{array}{*{20}c}
\vec {\sigma} \hfill & 0 \hfill \\
 0 \hfill & \vec {\sigma}\hfill \\
\end{array}
 \right)
\end{eqnarray}

\begin{equation}
\label{eq85}
\vec{J}=\vec{L} + \frac{1}{2}\vec{S}
\end{equation}

It straightforward to show that the Hamiltonian (\ref{eq80}) of
eqn. (\ref{eq81}) commutes with the operators $\vec {J}^ 2$ and
$J_{z}$ (the z- projections must be x or y). Using these
symmetries and the condition that the nonlinearity is of Kerr
type, we can solve the NDE equations (\ref{eq81}) by a separation
of variables technique.  We look for solutions in the form:

\begin{eqnarray}
\label{eq86}
 \Psi _{1} =  a\left(r\right)\Omega _{jlM} \\
 \Psi _{2} =  a\left(r\right)\Omega _{jlM} \\
 \Psi _{3} = ib\left(r\right)\Omega _{jl^{'}M} \\
 \Psi _{4} = ib\left(r\right)\Omega _{jl^{'}M},
\end{eqnarray}
where $\Omega _{jlm}$ is the spherical spinor, $l + l^{'} = 1$,
and $a\left(r\right)$ and $b\left({r}\right)$ are arbitrary radial
functions. Using the symmetries of (\ref{eq81}) and the fact, that
the nonlinear parts depend on $r$, we separate variables and
obtain the following system of equations for the radial part:

\begin{eqnarray}
\label{eq87}
\frac{\partial a\left(r\right)}{\partial r} +
\frac{1+\chi}{r}a\left(r\right) = -\nu_2 b\left(r\right)
\end{eqnarray}

\begin{eqnarray}
\label{eq88}
\frac{\partial b\left(r\right)}{\partial r} +
\frac{1-\chi}{r}b\left(r\right) =
\nu_1a\left(r\right)+\gamma{\left|a\left(r\right)\right|}^2a\left(r\right),
\end{eqnarray}
where
\begin{eqnarray}
\chi = l\left({l + 1}\right)-j\left({j+1}\right)-1/4.
\end{eqnarray}
Exclude $b(r)$ from the system (\ref{eq87})-(\ref{eq88}), we
obtained the next equation for $a(r)$:

\begin{eqnarray}
\label{eq87aa} \nu_1\nu_2a(r)+ \frac{\partial^2a}{\partial r^2}+
\frac{2}{r}\frac{\partial a}{\partial r} -
\frac{(1+\chi)\chi}{r^2}a +\nu_2\gamma{\left|a\right|}^2a=0.
\end{eqnarray}

Formally this equation admit exact "de Broglie" soliton solutions
for arbitrary number of $\chi$. But as we remember our solutions
are limited by the conditions (\ref{eq82}) the nonlinear part to
depends only on the radial components. The condition (\ref{eq82})
for a number $\chi\geq1$ can be fulfilled also for a higher number
of field on different frequencies. This case including also the
parametric processes. The case of one carrying frequencies
correspond to localized solutions with $\chi=1$ and angular
components $l=1$ and $j=\pm1/2$. In this case the system
(\ref{eq87})-(\ref{eq88}) becomes:

\begin{eqnarray}
\label{eq89}
\frac{\partial a\left(r\right)}{\partial r} +
\frac{2}{r}a\left(r\right)=-\nu_2b\left(r\right)
\end{eqnarray}

\begin{eqnarray}
\label{eq90}
\frac{\partial b\left(r\right)}{\partial r} =
\nu_1a\left(r\right)+\gamma{\left|a\left(r\right)\right|}^2a\left(r\right).
\end{eqnarray}

As was shown above, this system has exact radial solutions of the form
(\ref{eq65})-(\ref{eq67}):

\section{Experimental conditions}

There are some differences between the nonlinear conditions for
localized solutions of the Vector Nonlinear Schrodinger Equation
(VNLS) and localized conditions for the Nonlinear Maxwell
Equations (NME). The nonlinear parameter for the VNLS is written
\cite{AK}:

\begin{equation}
\label{eq75} \gamma_{vnls}=k_0^2r_0^2n_2{\left|A_0\right|}^2 = 1.
\end{equation}
For localized solutions in optical region the constant $\alpha ^2$
ranges from:

\begin{equation}
\label{eq76} \alpha ^2 = k_0^2 r_0^2 \approx 10^4 - 10^6,
\end{equation}
which corresponds to a required nonlinear refractive index change
of order of:

\begin{equation}
\label{eq77} n_2{\left|A_0\right|}^2\approx 10^{-4}-10^{-6}.
\end{equation}
On the other hand, the nonlinear condition for localized solutions
of the normalized NDE (\ref{eq38})-(\ref{eq41}) is:

\begin{equation}
\label{eq78}
\gamma_{NME}=\nu_2\gamma_1\approx(\delta\Delta\alpha-\alpha_2)
k_1r_0n_2{\left|A_0\right|}^2=1 .
\end{equation}
For the typical value of the constant $\nu_{2}\approx1$ the next
required nonlinear refractive index change appears:

\begin{equation}
\label{eq79} n_2{\left|A_0\right|}^2\approx 10^{-2}-10^{-3}.
\end{equation}
Another difference between these two cases is that the solutions
of the VNLS are comprised of linearly polarized components and
that the dispersion of the nonlinear medium plays an important
role. This leads to the fact that vortex solutions of the VNLS may
be observed only in special dispersion regions of nonlinear media.
The vortices of the NDE are a combination of linear and circular
polarization components, and do not have this marked dependence on
the dispersion and may be observed in the transparency region of a
nonlinear media.

\section {Vortex solutions in a nonlinear Kerr type media without dispersion}

      The case without linear dispersion of the electrical and magnetic susceptibility will
 correspond to $\chi^{(1)}=const$ and $\eta^{(1)}=const$ . We suppose now that the
 amplitude  functions do not depend from time and will look for 3D+1
 monochromatic electric and magnet field of kind:
 \begin{eqnarray}
\label{d1} \vec E(x,y,z,t)=\vec
M(x,y,z)\exp{\left(i\omega_0t\right)},
\end{eqnarray}

\begin{eqnarray}
\label{d2} \vec H(x,y,z,t)=\vec
N(x,y,z)\exp{\left(i\omega_0t\right)},
\end{eqnarray}
where $\vec M$ $\vec N$ and $\omega_0$ are the amplitudes of the
electric, magnetic fields and the optical frequency respectively.

Substituting the relations (\ref{d1})-(\ref{d1}) one obtain the
next amplitude equations in the case of dispersionless media:

\begin{eqnarray}
\label{h4} \nabla \times \vec {M} = i\alpha _{2} \vec {N},
\end{eqnarray}

\begin{equation}
\label{h5} \nabla \times \vec {N} = i\alpha _{1} \vec {M} +
\gamma\left(\vec M\cdot\vec M^{\ast}\right)\vec M,
\end{equation}

\begin{equation}
\label{h6} \nabla \cdot \vec {M} = 0
\end{equation}

\begin{equation}
\label{h7} \nabla \cdot \vec {N} = 0
\end{equation}

Again the electric and magnetic fields are represented as a sum of
a linear polarization component and a circular polarized one.
Substituting:

\begin{equation}
\label{h8} \psi _{1} = iM_l = iM_z
\end{equation}

\begin{eqnarray}
\label{h9} \psi _{2} = M_{c} = iM_{x} - M_{y}
\end{eqnarray}

\begin{eqnarray}
\label{h10} \psi_{3} = N_l=-iN_{z}
\end{eqnarray}

\begin{equation}
\label{h11} \psi_{4}=N_{c}=-N_{x} - iN_{y},
\end{equation}
into the nonlinear system (\ref{h4})-(\ref{h7}), we obtain the
same kind of stationary nonlinear Dirac system of equations (NDE)
as (\ref{eq52})-(\ref{eq55}) :

\begin{eqnarray}
\label{h12} \left(\frac{\partial}{\partial
x}-i\frac{\partial}{\partial y}\right) \Psi_4 +
\frac{\partial}{\partial z}\Psi_3 = -
i\left(\kappa_1+\zeta\sum\limits_{i =
1}^{2}{{\left|\Psi_i\right|}^2}\right) \Psi_1
\end{eqnarray}

\begin{equation}
\label{h13} \left(\frac{\partial}{\partial
x}+i\frac{\partial}{\partial y}\right) \Psi_3 -
\frac{\partial}{\partial z}\Psi_4 = -
i\left(\kappa_1+\zeta\sum\limits_{i =
1}^{2}{{\left|\Psi_i\right|}^2}\right) \Psi_2
\end{equation}

\begin{equation}
\label{h14} \left(\frac{\partial}{\partial x} -
i\frac{\partial}{\partial y}\right) \Psi_2 +
\frac{\partial}{\partial z}\Psi_1 = - i\kappa_2\Psi_3
\end{equation}

\begin{eqnarray}
\label{h15} \left(\frac{\partial}{\partial x} +
i\frac{\partial}{\partial y}\right) \Psi_1 -
\frac{\partial}{\partial z}\Psi_2 = - i\kappa_2\Psi_4,
\end{eqnarray}
but with different coefficients:
$\kappa_1=\frac{\omega_0\varepsilon_0}{c}$;
$\kappa_2=\frac{\omega_0\mu_0}{c}$;
$\zeta=\frac{4\pi\omega_0\chi{(3)}}{c}$. Naturally, the NDE
(\ref{h12})-(\ref{h15}) admit the same kind of solutions
(\ref{eq66})-(\ref{eq67})

\section {Conditions for finiteness of the energy of the vortex solutions}

 In this section we will investigate the localized energy in the
 both cases; with and without of linear dispersion. This
 corresponds
 to two kind of optical vortices; with and without spectral
 bandwidth.
 In a dielectric media without dispersion the expression for the linear part
 of energy density is:

\begin{eqnarray}
\label{e1} W_{lin} = \frac {1}{8\pi}
\left(\varepsilon\left|\vec E\right|^2+
  \mu\left|\vec H\right|^2\right),
\end{eqnarray}
where $\varepsilon$ and $\mu$ are constants. Substituting the
vortex solutions (\ref{eq66})-(\ref{eq67}) in (\ref{e1}) and
integrating over 3D space one obtain that quasi-monochromatic
vortices in a dielectric media without dispersion admit
\textit{infinite} energy. Now we come to the case of slowly
varying amplitude approximation and to calculation of  energy of
the vortex solutions of a media with linear electric and magnet
dispersion. To prove the finiteness of energy of the vortex
solutions (\ref{eq66})-(\ref{eq67})  we start with the equations
for averaged in time balance of energy density of electrical and
magnet field \cite{LAN}:

\begin{eqnarray}
\label{b3} \left\langle\frac{\partial W}{\partial t}
\right\rangle= \frac {1}{16\pi}\left(\vec E\cdot\frac{\partial
\vec D^{*}}{\partial t} + \vec E^{*}\cdot\frac{\partial \vec
D}{\partial t} +
  \vec B\cdot\frac{\partial \vec B^{*}}{\partial t} +
  \vec B^{*}\cdot\frac{\partial \vec B}{\partial t}\right),
\end{eqnarray}
where $\vec D=\vec P_{lin}+4\pi\vec P_{nlin}$ is a sum of the
linear induction and the nonlinear induction of the electrical
field. The calculations of the averaged energy of the optical
waves in a dispersive media are worked out considering the first
order of slowly varying amplitude approximation of electrical
induction (the same as in the NDE). The result comes to the old
result of Brillouin 1921 for energy density of electrical field:

\begin{eqnarray}
\label{b4} \left\langle W_{lin} \right\rangle= \frac{1}{8\pi}
\left(\frac{\partial\left(\omega\hat{\varepsilon}_0\right)}
{\partial \omega}{\left|\vec A\right|}^2+
\frac{\partial\left(\omega\hat{\mu}_0\right)}{\partial \omega}
{\left|\vec C\right|}^2\right).
\end{eqnarray}
The conditions of electric constant:

\begin{eqnarray}
\label{e0} \hat{\varepsilon}_0>0;
 \frac{\partial\left(\omega\hat{\varepsilon}_0\right)}
{\partial \omega}>0
\end{eqnarray}
is fulfilled in most of the dielectrics. The conditions of the
magnetic constant can be present in two cases. In the first one:

\begin{eqnarray}
\label{m01}\hat{\mu}_0>0;
 \frac{\partial\left(\omega\hat{\mu}_0\right)}
{\partial \omega}>0.
\end{eqnarray}
This condition corresponds to the case, when there are not
included any magnet relaxation processes. To include these
processes is possible only for ultrashort optical pulses with time
duration of dimension of local structure of one paramagnetic
media. The real part of magnet susceptibility in this case is
symmetric function in respect to the paramagnetic resonance and
the derivative in respect to the frequency is always negative.

\begin{eqnarray}
\label{m03}\hat{\mu}_0>0;
 \frac{\partial\hat{\mu}_0(\omega)}{\partial \omega}<0.
\end{eqnarray}
Under appropriate conditions this leads to:

\begin{eqnarray}
\label{m04}\hat{\mu}_0>0;
 \frac{\partial\left(\omega\hat{\mu}_0(\omega)\right)}
{\partial \omega}<0.
\end{eqnarray}

We can find spectral region and dispersion parameters when the
next condition will be satisfied:

\begin{eqnarray}
\label{m06} \frac{\partial\left(\omega\hat{\mu}_0\right)}
{\partial \omega}=-
 \frac{\partial\left(\omega\hat{\varepsilon}_0\right)}
{\partial \omega}.
\end{eqnarray}

Thus, the linear (infinity) part of energy density is zero. The
nonlinear part of averaged energy density is expressed in
\cite{SHEN} and for the vortex solutions (\ref{eq66})-(\ref{eq67})
becomes:

\begin{eqnarray}
\label{b10} \left\langle W_{nlin}\right\rangle =
n_2{\left|\Psi\right|}^2\Re(\Psi^2)+
 \omega_0\frac{\partial n_2}{\partial\omega_0}
 {\left|\Psi\right|}^2\Re(\Psi^2).
\end{eqnarray}
These results give the conditions for the finiteness of energy of
the vortices. Integrating $W_{nlin}$ in the 3D space we obtain a
finite value proportional to the main frequency $\omega_0$.

\section { Spatial phase functions, Poynting vector and flow of
energy}

The kind of the phase functions which satisfied
(\ref{phas1})-(\ref{phas2}) is determinate  for vortex solutions
(\ref{eq72}) with spin $j=1/2$. Using again the relations between
the spinors of NDE and the amplitude functions
(\ref{eq48})-(\ref{eq51}) we have:

\begin{eqnarray}
\label{am1} F_x=\frac{\psi _{2} -\psi_{2}^{*}}{2i};\
F_y=\frac{\psi_{2}+\psi_{2}^{*}}{2};\
F_z=\frac{\psi_{1}-\psi_{1}^{*}}{2i},
\end{eqnarray}

\begin{eqnarray}
\label{am4} G_x=-\frac{\psi _{4}+\psi_{4}^{*}}{2};\
G_y=-\frac{\psi_{4}-\psi_{4}^{*}}{2i};\
G_z=-\frac{\psi_{3}-\psi_{3}^{*}}{2i}.
\end{eqnarray}
Substituting the solutions (\ref{eq72}) with spin $j=1/2$  in
(\ref{am1})-(\ref{am4}) for arbitrary real $a(r)$ and $b(r)$ we
obtain:

\begin{eqnarray}
\label{x1} F_x=-a(r)\frac{x}{r};\ F_y=-a(r)\frac{y}{r};\
F_z=-a(r)\frac{z}{r},
\end{eqnarray}

\begin{eqnarray}
\label{x4} G_x=0;\ G_y=0;\ G_z=b(r).
\end{eqnarray}
We rewrite again the conditions for the spatial phase functions:

\begin{eqnarray}
\label{phase1} \nabla g\times \vec {F} = 0;\ \nabla q\times
\vec{G} = 0.
\end{eqnarray}
These relations for solutions of kind (\ref{x1})-(\ref{x4}) are
satisfied only when:

\begin{eqnarray}
\label{kr} g=k_0r\ or \ g=k_0f(r),
\end{eqnarray}
and

\begin{eqnarray}
\label{kz} q=k_0z\ or\ q=k_0f(z),
\end{eqnarray}
where $k_0$ is the carrying wave number. The spatial phase
functions of kind $g=k_0r$ and $q=k_0z$ correspond to spectral
limited pulses which satisfied additional relations:

\begin{eqnarray}
\label{delkr} \triangle k \triangle r= const.
\end{eqnarray}
The spatial phase functions of kind $g=k_0f(r)$ and $q=k_0f(z)$
correspond to phase modulated pulses and for them the relations
(\ref{delkr}) is not satisfied.  The Poynting vector can be
expressed by the amplitude functions of the electrical and magnet
field:

\begin{eqnarray}
\label{Poyn}
\vec{S}=\vec{E}(x,y,z,t)\times\vec{H}(x,y,z,t)=\nonumber\\
\exp\left(i\left(W(t)-K(r)\right)\right)\vec{F}(x,y,z,t)\times\vec{G}(x,y,z,t)
\end{eqnarray}

where $W$ and $K$ are scalar phase functions. Substituting the
solutions with spin $j=1/2$ in above expression we find that:
\begin{eqnarray}
\label{Pn12} \vec{S}=
\exp\left(i\left(W(t)-K(r)\right)\right)\left(-a(r)b(r)\frac{y}{r};\
a(r)b(r)\frac{x}{r};\ 0\right).
\end{eqnarray}

We see that the Poynting vector $\vec{S}$ is one circulation
vector for solutions with spin $j=1/2$ and its divergency is zero:

\begin{eqnarray}
\label{divS} \nabla\cdot\vec{S}=0.
\end{eqnarray}
The relation (\ref{divS}) determine that the energy flow through
arbitrary closed surface around our vortex solutions with spin
$j=1/2$ is zero. The relation (\ref{Pn12}) show that flow of
energy of our solutions circulate in $x,y$ plane. Now we can
generalize the above results for solutions with spin $j=1/2$: The
vortex solutions with spin $j=1/2$ without external fields are
immovable and electromagnetic energy oscillated in $x,y$ plane.
The electrical field oscillating spherically, in 'r' direction,
while the magnet field oscillating in z direction. In the same way
was calculated the Poynting vector for solutions with spin
$j=-1/2$. For them we obtain that $\nabla\cdot\vec{S}\neq0$ and we
expect that they are not stable. One exact investigation of
stability request investigation also the perturbation of the
Poynting theorem and will discussed in one next paper.

\section{ Conclusion}

In this paper we derive a set of Nonlinear Maxwell amplitude
Equations (NME) for nonlinear optical media with and without
dispersion of the electric and magnetic susceptibility. We have
shown that in cases of linear and circularly polarized components
of the electric and magnetic fields, the NME reduces to the
Nonlinear Dirac system of equations (NDE). The equations are
represented in a spinor form. Using the method of separation of
variables, exact vortex solutions for both cases have been
obtained. The optical vortex solutions admit classical orbital
momentum $l = 1$ and classical own momentum $j = \pm1 / 2$. Here I
would like to explain more abbout the differences between
solutions, obtained by separation of variables of the usual linear
Dirac equation with potential depending only on 'r' (for example
hydrogen atom) and solutions of the NDE. In case of linear Dirac
equations with potential to higher order spherical spinors
correspond higher order of radial spherical Bessel functions. On
the other hand for NDE to higher order of spherical spinors
($\ell=1,2,..$) correspond higher number of field components and
higher value of localized energy. For all radial solutions of NDE
the zero spherical radial Bessel function $\frac{\sin\alpha r}{r}$
are valid. The optical vortices in a media without dispersion
admit infinite energy integral. The energy integral of the vortex
solutions is finite only in some special cases of paramagnetic
media with suitable conditions on linear electric and magnet
dispersion . Using the Poynting vector for solutions with spin
$j=1/2$ we find that the energy flow through arbitrary closed
surface around our vortex solutions is zero and the localized
energy of our solutions circulate in $x,y$ plane. Other important
result is, that the vortex solutions with spin $j=1/2$ without
external fields are immovable. The electrical field in the
vortices oscillating spherically, while the magnet field
oscillating in z direction. The initial investigations on
stability of these solutions show the following: While the
vortices with spin $j=1/2$ are stable, the vortices with opposite
spin (charge) $j=-1/2$ are not.

All of the above results will  be discussed later in relation with
nonlinear field theory. However one direct approach to the plasma
physics can be found. We consider the old idea presented by
Gaponov and Miller in \cite{GM}: Confinement of charged particles
in potentials of high frequency electromagnetic fields. In our
article are presented solutions inspired by this idea. The optical
vortices are of such type of high frequency fields with zero
intensity in the centrum. In a cold plasma it is possible to
obtain  trapping of ions in the field of vortices. Thus,
conditions for ruled nuclear fusion can be reached.

\end{document}